\documentclass[lettersize,journal]{IEEEtran}
\IEEEoverridecommandlockouts

\usepackage[numbers]{natbib}

\usepackage{amsthm}
\newtheorem{definition}{Definition}
\newtheorem{example}{Example}
\usepackage{amsmath,amssymb,amsfonts}
\usepackage{algorithmic}
\usepackage{graphicx}
\usepackage{subfigure}
\usepackage{textcomp}
\usepackage{newtxtext}
\usepackage{xcolor}

\usepackage{bm}
\usepackage[linesnumbered,ruled,vlined]{algorithm2e}
\usepackage{dashbox}
\usepackage{url}
\usepackage{multirow}
\usepackage{booktabs}
\usepackage{flushend}
\usepackage{balance}
\usepackage{threeparttable}
\usepackage{comment}
\usepackage{balance}
\usepackage{pifont}

\usepackage{makecell}

\usepackage{orcidlink}

\newcommand{\eg}{\emph{e.g.,}\xspace}
\newcommand{\ie}{\emph{i.e.,}\xspace}

\newcommand{\diversitysmt}{diversity-aware SMT algorithm\xspace}
\newcommand{\smtsampler}{\emph{SMTSampler}\xspace}
\newcommand{\csb}{\emph{csb}\xspace}
\newcommand{\guidedsampler}{\emph{GuidedSampler}\xspace}
\newcommand{\bitwuzla}{\emph{Bitwuzla}\xspace}

\newcommand{\thismethod}{\emph{PanSampler}\xspace}
\newcommand{\score}{AST-guided scoring function\xspace}
\newcommand{\postopt}{post-sampling optimization technology\xspace}

\newcommand{\altone}{\emph{Alt-1}\xspace}
\newcommand{\alttwo}{\emph{Alt-2}\xspace}
\newcommand{\altthree}{\emph{Alt-3}\xspace}

\begin{document}

\title{Towards Comprehensive Sampling of SMT Solutions
\thanks{This work was supported in part by the National Key Research and Development Program of China under Grant 2023YFB3307503, in part by the National Natural Science Foundation of China under Grants 62202025 and 62302528, in part by Beijing Natural Science Foundation under Grant L241050, in part by the Young Elite Scientist Sponsorship Program by CAST under Grant YESS20230566, in part by CCF-Huawei Populus Grove Fund under Grant CCF-HuaweiFM2024005,
and in part by the Fundamental Research Fund Project of Beihang University. \textit{(Corresponding author: Chuan Luo.)}}
}

\author{Shuangyu~Lyu~\orcidlink{0009-0003-2040-9428},
        Chuan~Luo~\orcidlink{0000-0001-5028-1064},
        Ruizhi~Shi~\orcidlink{0009-0008-0520-6231},
        Wei~Wu~\orcidlink{0000-0002-0975-4613},
        Chanjuan~Liu~\orcidlink{0000-0002-8132-2380},
        Chunming~Hu~\orcidlink{0000-0003-3473-9703}
\thanks{Shuangyu Lyu, Chuan Luo, Ruizhi Shi and Chunming Hu are with the School of Software, Beihang University, Beijing 100191, China (e-mail: shuangyulyu@buaa.edu.cn; chuanluo@buaa.edu.cn; shiruizhi@buaa.edu.cn;
hucm@buaa.edu.cn).}

\thanks{Wei Wu is with the School of Computer Science and Engineering, Central South University, Changsha 410083, China (e-mail: wei.wu@csu.edu.cn).}

\thanks{Chanjuan Liu is with the School of Computer Science and Technology, Dalian University of Technology, Dalian 116024, China (e-mail: chanjuanliu@dlut.edu.cn).}
}

\markboth{}%
{Lyu \MakeLowercase{\textit{et al.}}: Towards Effective Sampling of SMT Solutions}

\maketitle

\begin{abstract}
This work focuses on effectively generating diverse solutions for satisfiability modulo theories (SMT) formulas, targeting the theories of bit-vectors, arrays, and uninterpreted functions, which is a critical task in software and hardware testing. Generating diverse SMT solutions helps uncover faults and detect safety violations during the verification and testing process, resulting in the SMT sampling problem, \ie{} constructing a small number of solutions while achieving comprehensive coverage of the constraint space. While high coverage is crucial for exploring system behaviors, reducing the number of solutions is of great importance, as excessive solutions increase testing time and resource usage, undermining efficiency. In this work, we introduce \thismethod, a novel SMT sampler that achieves high coverage with a small number of solutions. \thismethod incorporates three novel techniques, \ie{} \diversitysmt, abstract syntax tree (AST)-guided scoring function and \postopt, which enhance its practical performance. It iteratively samples solutions, evaluates candidates, and employs local search to refine solutions, ensuring high coverage with a small number of samples.
Extensive experiments on practical benchmarks demonstrate that \thismethod exhibits a significantly stronger capability to reach high target coverage, while requiring fewer solutions than state-of-the-art samplers to achieve the same coverage level.
Furthermore, our empirical evaluation on practical subjects, which are collected from real-world software systems, shows that \thismethod achieves higher fault detection capability and reduces the number of required test cases from 32.6\% to 76.4\% to reach the same fault detection effectiveness, leading to a substantial improvement in testing efficiency.
Our evaluations also validate the effectiveness of \thismethod's core techniques.
\thismethod represents a significant advancement in SMT sampling, contributing to the cost reduction of software testing and hardware verification.
\end{abstract}

\section{Introduction}
\label{sec:introduction}

Satisfiability modulo theories (SMT) plays a fundamental role in testing and verification of software and hardware systems \cite{MeGASampler}.
In software verification \cite{ClaEtAl03,GurEtAl15,HeiEtAl13}, SMT formulas are employed to represent safety requirements derived from the underlying code.
A prime example is symbolic execution, where constraints from the code are encoded and solved to generate inputs that explore uncovered paths \cite{JFSampler}.
Similarly, in constrained-random verification (CRV) \cite{NavEtAl06}, SMT formulas encode the functional model and test scenarios of the hardware design.
Among the various SMT theories, bit-vectors, arrays, and uninterpreted functions are particularly important and widely applied in practice. 
This work focuses on these theories, since they are fundamental to many SMT-based applications in software and hardware verification \cite{SMTSampler,GuidedSampler,LuoEtAl22}.
While SMT solvers, which have been extensively studied
\cite{BarEtAl11,BarEtAl22,CimEtAl13,DutEtAl14,MouEtAl08,Bitwuzla,BruEtAl09},
can determine the satisfiability of a formula and provide a single solution, certain applications require the generation of diverse solutions.
For instance, diverse solutions can accelerate symbolic execution and reap the benefits of random testing \cite{QuickSampler}.
In the context of software verification, these solutions help identify inputs that trigger safety violations or faulty program behavior.
In CRV, they increase the likelihood of uncovering hardware faults.
Generating diverse solutions thus enables the detection of a wider range of faults in both software and hardware systems, providing valuable insights for system analysis and issue identification \cite{ClaEtAl03,HeiEtAl13}.

In software and hardware testing,
diverse SMT solutions are essential for ensuring that test cases and stimuli cover a broad range of scenarios, thereby enhancing the capability of detecting hidden faults or violations.
However, while diversity is crucial, it must be balanced with testing efficiency, as executing a large number of test cases can be computationally expensive \cite{LuoEtAl22}.
Generating an excessively large number of
test cases or stimuli
significantly increases the required resource and poses technical challenges, especially in large-scale systems or environments with tight development schedules \cite{NoeEtAl20}.
In software testing, executing a large test suite slows down the development cycle \cite{NoeEtAl20}, while in
CRV,
simulating a vast number of stimuli requires substantial computational power and storage.
By generating a small yet diverse set of SMT solutions, practitioners can optimize fault detection while reducing the overhead associated with
practical verification.
The importance of reducing the number of SMT solutions has been emphasized by extensive studies
\cite{LuoEtAl22,CruEtAl19,HarEtAl93,NoeEtAl20,CheEtAl20,JehEtAl23}.
This leads to the SMT sampling problem, which aims to build small-sized solution sets for SMT formulas while achieving high coverage of the constraint space \cite{SMTSampler,MeGASampler}.

As mentioned,
a diverse solution set with high constraint space coverage is crucial to covering diverse scenarios and detecting hidden faults.
To measure the constraint space coverage of solution sets,
several
studies have proposed using the uniformity of the solution distribution as a proxy for constraint space coverage \cite{QuickSampler,HerEtAl20}.
However, accurately assessing distribution uniformity requires sampling a prohibitively large number of solutions, which is impractical for large-scale SMT formulas.
Furthermore, recent empirical work \cite{SMTSampler,LuoEtAl21b} has demonstrated that uniform sampling is neither consistently effective for detecting faults in
the verification and testing process,
nor sufficient for achieving the optimal coverage of constraint space.
To address these limitations, a recent study \cite{SMTSampler} has introduced the abstract syntax tree (AST)-coverage of an SMT formula (formally defined in Section \ref{sec:preliminaries}) as an effective measure of constraint space coverage.
This metric captures the range of syntactic structures and decision paths exercised within the formula, making it a reliable proxy for constraint space coverage \cite{SMTSampler,MeGASampler}.
Also, a solution set with high AST-coverage leads to improved fault detection capability \cite{SMTSampler,MeGASampler}.
Consequently, subsequent SMT sampling research has adopted this metric as a standard evaluation criterion \cite{SMTSampler,MeGASampler,JFSampler} and a key optimization objective in algorithm development \cite{JFSampler}.

Compared to the problem of boolean satisfiability (SAT) sampling that has been extensively studied \cite{KitEtAL07,KitEtAL10,ChaEtAl14,ChaEtAl15,ErmEtAl13,NadEtAl11,GolEtAl21,QuickSampler}, the SMT sampling problem has received much less attention \cite{SMTSampler,GuidedSampler}.
Although SMT formulas can be converted into SAT formulas using eager bit-blasting techniques
in certain theories (\eg{} bit-vectors),
enabling the application of SAT sampling methods, such as the recent method \csb \cite{csb}, this conversion usually incurs a loss of higher-level structure in the formula.
The diversity of sampled solutions is thus reduced \cite{SMTSampler}.
This issue is further supported by empirical evidence from prior work \cite{SMTSampler} and our experimental results in Section \ref{sec:experimental_results}.
State-of-the-art SMT samplers \cite{SMTSampler,GuidedSampler}, which utilize the mutation combination technique \cite{SMTSampler} to refine solutions, have successfully improved sampling efficiency.
However, this enhancement comes at the cost of reduced diversity in the generated samples.
In the theory of linear integer arithmetic, a model-guided method has been proposed \cite{MeGASampler}, but this method is limited by the properties of linear arithmetic operations and can only handle such operations.
Handling complex nonlinear constraints in theories such as bit-vectors, arrays, and 
uninterpreted
functions remains beyond its capabilities.
For the floating-point theory, an SMT sampler has been developed using a specialized SMT solver tailored for floating-point constraints \cite{JFSampler}.
While this approach improves
floating-point sampling,
its applicability to other theories remains limited.

In practice, existing SMT samplers face significant scalability challenge \cite{ChaEtAl14,ChaEtAl15,KitEtAL10}, \ie{} they cannot effectively deal with large-scale SMT formulas.
In particular, when handling large-scale SMT formulas, existing SMT samplers would produce solutions that concentrate in a specific region of the constraint space, thereby failing to cover the entire space adequately and making it difficult to reach the target coverage. 
Moreover, existing samplers
do
not explicitly optimize the number of sampled solutions required to achieve the target coverage \cite{SMTSampler,GuidedSampler,MeGASampler}, resulting in a large solution set that adversely impacts software testing and hardware verification, as discussed before.

In this work, we introduce \thismethod, which is a novel and effective SMT sampler, targeting the theories of bit-vectors, arrays, and uninterpreted functions.
Compared to existing methods, during its sampling process,
\thismethod explicitly minimizes the number of solutions required to achieve the target coverage.
In addition, we propose
three novel techniques, \ie{} \diversitysmt, AST-guided scoring function and the \postopt,
enabling \thismethod to obtain high
target
coverage with fewer solutions, thereby significantly improving both testing efficiency and effectiveness in practice.
Particularly, \thismethod iteratively samples solutions until a set that achieves the target coverage is built.
In each iteration, \thismethod consists of three phases, \ie{} sampling phase, evaluation phase and optimization phase.
During the sampling phase, \thismethod samples multiple candidate solutions that are likely to enhance coverage, forming a candidate set.
In the evaluation phase, \thismethod evaluates each solution in the candidate set and selects the most suitable one.
In the optimization phase, \thismethod uses the selected solution as a starting point for local search, attempting to find better solutions.
Through this way, \thismethod is able to construct a small solution set with high coverage.
To evaluate \thismethod, we conducted extensive experiments on practical benchmarks from the theories of bit-vectors, arrays, and uninterpreted functions (\ie{} from the logics of QF\_BV,  QF\_ABV, and QF\_AUFBV), which have been widely studied \cite{SMTSampler,GuidedSampler,YaoEtAl20,Bitwuzla,BruEtAl09}.
Our experimental results show that, compared to existing SMT samplers, \thismethod demonstrates a stronger ability to reach high target coverage while requiring fewer solutions to achieve the same coverage level.
Furthermore, our empirical evaluation on practical subjects, which are collected from real-world software systems, confirms that \thismethod enhances fault detection capability and, to achieve the same fault detection effectiveness, reduces the number of required test cases by between 32.6\% and 76.4\%, thus substantially improving practical software testing efficiency.

Our main contributions
are summarized as follows.
\begin{itemize}

    \item We introduce a new and effective SMT sampler dubbed \thismethod, targeting the theories of bit-vectors, arrays, and uninterpreted functions.

    \item We propose three novel techniques, \ie{} \diversitysmt, \score and \postopt, to
    enhance the practical performance of \thismethod.

    \item We conducted extensive experiments to demonstrate the superiority of \thismethod over existing SMT samplers,
    both in sampling performance and in enhancing practical software testing,
    advancing the state of the art in SMT sampling.
    
\end{itemize}

\section{Preliminaries}
\label{sec:preliminaries}

\subsection{Definitions and Notations Related to SMT Sampling}
\label{sec:definitions_and_notations}

\textbf{SAT Formula.}
A SAT formula
is a formula constructed from a set of Boolean variables, where the variables are combined using propositional logic operators (\eg{} $\land$, $\lor$, $\neg$).

\textbf{SMT Formula.}
An SMT formula is a logic formula that is interpreted with respect to a background theory \cite{BarEtAl18}.
Unlike a purely propositional SAT formula, an SMT formula incorporates symbols and functions from specific theories, such as bit-vectors, arrays, and uninterpreted functions.
This extension enables the precise modeling of a broader range of complex problems.
In the experimental sections (\ie{} Section \ref{sec:experimental_design} and Section \ref{sec:experimental_results}), SMT formulas are also referred to as SMT benchmarks.

\textbf{Theory of SMT Formula.}
The theories
of SMT formula constrains the domain of the variables and the set of the operators.
In this work, we focus on the theories of bit-vectors, arrays, and uninterpreted functions, which are described below.
That is, the SMT formulas considered in this work are quantifier-free formulas over the theories of bit-vectors, bit-vector arrays, and uninterpreted functions.
For an SMT formula $\varphi$, $V(\varphi)$ denotes the set of all variables in $\varphi$.

\begin{itemize}
    \item \textbf{Bit-vector.}
    A bit-vector represents a fixed-size sequence of bits, which are of binary values (either 0 or 1).
    The bit-vector theory contains operators including bitwise operators (\eg{} bitwise \texttt{shift}, bitwise \texttt{and}) and
    arithmetic operators
    (such as $+$, $\times$, $<$).
    A bit-vector of size $m$ is denoted as $BV_m$, where each bit in the vector is regarded as a boolean variable.
    Given a variable $x$ of type $BV_m$, the $i$-th lowest bit of $x$ is denoted as $x(i)$, where $0 \leq i \leq m - 1$,
    and the lowest bit corresponds to the least significant bit, indexed as 0.

    \item \textbf{Array.}
    An array is a mapping, where each element can be accessed using an index.
    The theory of arrays provides two primary operations: $\texttt{select}$ and $\texttt{store}$, which are used to retrieve and modify elements, respectively.
    A bit-vector array is a special type of array where each element is a bit-vector, and all elements have the same size.

    \item \textbf{Uninterpreted Functions.}
    Uninterpreted functions are symbolic functions without concrete definitions, represented as $(f\ x_1\ x_2\ \cdots\ x_n)$.
    For example, $f : BV_1 \times BV_2 \rightarrow BV_2$ defines a function $f$ that takes a bit-vector of size 1 and a bit-vector of size 2 as input and returns a bit-vector of size 2.
    These functions are treated symbolically during solving.   
\end{itemize}

\textbf{Total Variable Bits.}
For an SMT formula $\varphi$ that involves bit-vectors, the total variable bits is defined as
the sum of the bit-widths of all variables in $V(\varphi)$.
For example, if $\varphi$ contains a bit-vector variable $x$ of size $m$ and a bit-vector variable $y$ of size $n$, then the total variables bits $\varphi$ is $m + n$.
The set of all bits of all variables in $V(\varphi)$ is denoted as $\mathit{VarBits(\varphi)}$, meaning that the total variable bits of $\varphi$ is $\mathit{|VarBits(\varphi)|}$.
This metric is crucial for assessing the complexity of an SMT formula.

\textbf{Theory Restriction.}
Given two theories $T$ and $T'$, we say that theory $T'$ is a restriction of theory $T$, denoted as $T' \subseteq T$, if and only if each valid formula $\varphi$ in $T'$ is also valid in $T$.

\textbf{Assignment.}
Given an SMT formula $\varphi$ with its set of variables $V(\varphi)$, an assignment of $\varphi$ is
a mapping $\sigma$ from each variable $x \in V(\varphi)$ to its assigned value, denoted as $\sigma[\![x]\!]$.
A solution (also known as satisfying assignment) of $\varphi$ is an assignment that makes formula $\varphi$ be satisfied.
For an SMT formula $\varphi$ in $T$ and a formula $\varphi'$ in $T'$, with $V(\varphi') \subseteq V(\varphi)$ and $T' \subseteq T$, if $\sigma$ is an assignment of $\varphi$, then $\sigma[\![\varphi']\!]$ denotes the value of $\varphi'$ under assignment $\sigma$.
$S(\varphi)$ denotes the set of $\varphi$'s all solutions.

\textbf{Abstract Syntax Tree (AST).}
The AST of an SMT formula $\varphi$ is a hierarchical representation of $\varphi$, where each node corresponds to a sub-expression or a term in the formula.

\textbf{AST-bit.}
An AST-bit is a tuple $(b,v)$, where $b$ is a bit of a boolean or bit-vector node in the AST, and $v$ is either 0 or 1.
For example, given a node $x$ of type $BV_3$, it can derive 6 AST-bits: $(x(0), 0)$, $(x(0), 1)$, $(x(1), 0)$, $(x(1), 1)$, $(x(2), 0)$ and $(x(2), 1)$.
An AST-bit $(b, v)$ is covered by a solution $\alpha$ if and only if $\alpha[\![b]\!] = v$.
An AST-bit is considered valid if and only if there exists at least one solution that
covers
it.
$\mathit{ASTBits(\Phi)}$ denotes the set of all AST-bits $(b,v)$ for which $v \in \{0,1\}$, derived from every bit of every node in the AST of $\Phi$.
$ValidBits(\varphi)$ is defined as the number of all valid AST-bits in the AST of $\varphi$.

\textbf{Cover Set of Solution.}
Given a solution $\alpha$, $CoverSet(\alpha, \varphi)$ denotes the set of $\varphi$'s all AST-bits that are covered by $\alpha$, \ie{} $CoverSet(\alpha, \varphi) = \{(b,v) \mid (b,v) \in ASTBits(\varphi) \land \alpha[\![b]\!] = v\}$.
Also, given a solution set $A$, $CoverSet(A, \varphi) = \underset{\alpha \in A}{\bigcup} CoverSet(\alpha, \varphi)$.

\textbf{AST-Coverage.}
Given an SMT formula $\varphi$ and its solution set $A$, notation $Coverage(A, \varphi)$ is the ratio of $|CoverSet(A, \varphi)|$ to $ValidBits(\varphi)$, which is also referred to as AST-coverage.
According to the literature \cite{SMTSampler,MeGASampler}, this metric effectively reflects a solution set's coverage of formula $\varphi$'s constraint space, and a solution set with higher AST-coverage usually has stronger fault detection capability in practice.

\textbf{Bit-Blasting.}
During SMT solving,
it is sometimes necessary to translate bit-vector SMT formulas into equisatisfiable SAT formulas.
The bit-blasting technique is widely employed by state-of-the-art bit-vector solvers to achieve this transformation \cite{Bitwuzla}.
Given a bit-vector SMT formula $\Phi$, the bit-blasting technique converts it into a corresponding SAT formula $\phi_b$.
A solution $\alpha_b$ of $\phi_b$ corresponds to a solution of $\Phi$, denoted as $\Phi_{\phi_b}(\alpha_b)$.
For any variable $x \in V(\Phi)$, each bit of $x$ (\ie{} $x(i)$), corresponds to a variable $x_b \in V(\phi_b)$.
This correspondence is represented as $blasted(x(i))=x_b$.

\subsection{Minimum Solution Set Coverage Problem}

As aforementioned, two critical metrics are used to evaluate the solution sets generated by SMT samplers, \ie{}
the
AST-coverage
and the cardinality of the solution set. 
The former determines fault detection capability \cite{SMTSampler,MeGASampler}, while the latter impacts the efficiency of practical verification and testing processes \cite{CruEtAl19,HarEtAl93,NoeEtAl20,CheEtAl20,JehEtAl23}.
In particular, constructing a solution set with a large cardinality incurs significant resource costs, which is a critical concern in software testing and hardware verification \cite{CruEtAl19,HarEtAl93,NoeEtAl20,CheEtAl20,JehEtAl23}.
Moreover, as discussed in Section \ref{sec:introduction},
AST-coverage is able to effectively reflect the coverage of constraint space \cite{SMTSampler,MeGASampler,JFSampler},
and a high AST-coverage solution set effectively captures diverse scenarios and reveals hidden faults.
Therefore, it is essential to generate as few solutions as possible while ensuring high
AST-coverage.

Based on the above discussion, the minimum solution set coverage problem is formally
defined as follows:
\begin{definition}
Given an SMT formula $\varphi$ and a real-valued threshold $r$ between 0 and 1 representing the target AST-coverage,
the minimum solution set coverage problem is to find a solution set $A$ of $\varphi$ such that the AST-coverage of $A$ is at least $r$, while minimizing the cardinality of $A$ (\ie{} $|A|$).
\end{definition}

Solving the minimum solution set coverage problem yields a solution set $A$ with strong fault detection capability and small cardinality, enhancing the effectiveness and efficiency of verification and testing in practice.
Therefore, effectively tackling this problem is of significant importance.
However, since the SMT sampling problem is computationally hard \cite{SMTSampler,MeGASampler}, our goal is to efficiently find a near-optimal solution set with a cardinality as small as possible.

\section{\thismethod: Our Proposed SMT Sampler}
\label{sec:our_proposed_algorithm}

This section presents
\thismethod, an effective SMT sampler for solving the minimum solution set coverage problem.

\subsection{Challenges and Potential Solutions}
\label{sec:challengs}

Before introducing \thismethod, we outline the challenges it aims to mitigate and discuss potential solutions.

\subsubsection{Scalability Challenge}

As discussed in Section \ref{sec:introduction}, minimizing the number of solutions to achieve the target AST-coverage is critical for efficient testing and verification processes \cite{CruEtAl19,HarEtAl93,NoeEtAl20,CheEtAl20,JehEtAl23}.
In particular, as modern software and hardware systems evolve, the SMT formulas used to model them for testing and verification have become increasingly large.
This trend leads to a growing demand for effective samplers capable of handling large-scale SMT formulas.
However, for such formulas,
the solution space can be excessively large.
Existing samplers typically perform random, uninformed searches guided by predefined approximation strategies.
Some randomize variable initializations \cite{GolEtAl21,QuickSampler}, while others enforce a specific target distribution, such as uniformity, across the solution set \cite{ChaEtAl14,ChaEtAl15,HenEtAl14}.
Few of them explicitly optimize for the AST-coverage.
A fuzzing-based sampler \cite{JFSampler} uses
ASTs
to guide sampling, but provides only coarse guidance, limiting its effectiveness.
When the SMT formula is small, achieving the high target AST-coverage is relatively easy, and these samplers may produce acceptable results.
However, for the large-scale formulas, they cause slow increases in AST-coverage as the number of solutions grows. Thus, achieving the target AST-coverage may fail or require a large number of solutions, impacting the efficiency of testing and verification.
Our experimental results in Section \ref{sec:experimental_results} demonstrate this point.

\textbf{Solution to Scalability Challenge:}
\thismethod adopts a focused, iterative process to construct the solution set, where each step is carefully designed to optimize AST-coverage and minimize the solution set size.
It applies novel \score to evaluate and select a promising solution firstly,
and then employs the \postopt to further improve the selected solution’s contribution to AST-coverage.
Therefore, each sampled solution is optimized to enhance the overall AST-coverage, yielding a small-sized solution set that meets the target AST-coverage.

\subsubsection{Diversity Challenge}
To minimizing the size of sampled solution sets,
an intuitive approach is to construct sets with high diversity.
However,
existing SMT solvers usually generate same or similar solutions across repeated invocations.
Some samplers rely on randomization to promote diversity, such as altering the random seed or modifying the initial search state \cite{GolEtAl21, QuickSampler}, resulting in limited guidance in
solution diversity.
Others impose additional constraints to further constrain the solution space and prevent identical solutions, but this also inherently increases solver complexity and reduces efficiency.
Moreover, all these approaches do not consider the current state of the already generated solutions.
In an iterative sampling process, to enhance the diversity of the final set, it is effective to ensure that each new solution is as distinct as possible from those already generated, rather than relying on purely random generation each time.

\textbf{Solution to Diversity Challenge:}
We designed a specialized SMT solver, \diversitysmt, tailored to this issue.
It generates new solutions that are distinct from the existing solution set.
\diversitysmt operates by generating solutions that deviate from the distribution of existing solutions, while avoiding the imposition of additional constraints that would significantly increase solver complexity.
By doing so,
the size of the sampled solution set is reduced.

\subsection{Overall Design of \emph{\thismethod}}
\label{sec:overall_design}

{
\SetAlCapFnt{\small}
\SetAlCapNameFnt{\small}
\SetAlFnt{\small}
\SetKw{Continue}{continue}
\SetKw{Break}{break}

\begin{algorithm} [t]
    \caption{Overall Design of \thismethod}
    \label{algo:overall_design}
    
    \KwIn{$\bm{\Phi}$: an SMT formula;
    \newline
    $\bm{r}$: the target AST-coverage\;}
    \KwOut{$\bm{A}$: the set of generated solutions\;}

    $A\leftarrow\varnothing$\;

    \While{termination condition is not met}
    {
        $C\leftarrow\varnothing$\;\label{algoline:sampling_begin}
        
        \For{$i \gets 1$ \textbf{to} $\lambda$}
        {
            $\alpha_i\leftarrow$\texttt{DiversitySMT}($\Phi$, $A$)\;
            
            $C\leftarrow C\cup\{\alpha_i\}$\;
        }\label{algoline:sampling_end}

        $Score\leftarrow\varnothing$\;\label{algoline:evaluation_begin}
        
        \ForEach{$\alpha$ in $C$}
        {
            $Score$($\alpha$)$\leftarrow$\texttt{ASTScrFunc}($\Phi$, $A$, $\alpha$)\;
        }

        $\alpha\leftarrow$ the solution that maximizes $Score$ in $C$\;\label{algoline:evaluation_end}
        
        $\alpha^*\leftarrow$\texttt{PostOpt}($\Phi$, $A$, $\alpha$)\;\label{algoline:optimization}
        
        $A\leftarrow A\cup\{\alpha^*\}$\;
    }

    \Return{$A$}\;
    
\end{algorithm}
}

As detailed in Algorithm \ref{algo:overall_design}, the input to \thismethod consists of an SMT formula $\Phi$ and a real number $r$ between 0 and 1, representing the target AST-coverage.
After receiving these inputs, \thismethod generates a solution set $A$ such that $Coverage(A, \Phi) \geq r$.
\thismethod employs an iterative approach, where a new solution is generated in each iteration.
Given that the goal is to reach the target AST-coverage $r$ with as few solutions as possible,
in each iteration, \thismethod searches for the optimal solution within the solution space $S(\Phi)$, maximizing its contribution to AST-coverage.

One intuitive approach is exhaustively traversing the entire solution space $S(\Phi)$ to select the optimal solution.
However, since $|S(\Phi)|$ can be extremely large, this is impractical.
Instead, \thismethod follows the framework introduced below.
First, random sampling is performed within $S(\Phi)$,
collecting solution locations
likely to contribute significantly
to AST-coverage, forming a candidate set $C$.
Next, we traverse the solutions in $C$, each representing a small region in the solution space, and select the optimal solution $\alpha$.
While $\alpha$ is optimal within C, it may not attain global optimality across the entire solution space.
Nevertheless, this selection provides a promising search direction, as a better solution might exist in the vicinity of $\alpha$.
Therefore, we then employ a local search within the vicinity of $\alpha$ to refine the solution, aiming to identify a potentially better solution.
Through the above approach,
each iteration focuses on identifying the solutions contributing most to the AST-coverage, leading to a small-sized solution set.

According to Algorithm \ref{algo:overall_design},
the process of generating each solution in every iteration is divided into three phases: sampling phase (Line \ref{algoline:sampling_begin} -- Line \ref{algoline:sampling_end} in Algorithm \ref{algo:overall_design}), evaluation phase (Line \ref{algoline:evaluation_begin} -- Line \ref{algoline:evaluation_end} in Algorithm \ref{algo:overall_design}), and optimization phase (Line \ref{algoline:optimization} in Algorithm \ref{algo:overall_design}).
In the sampling phase, \thismethod samples a candidate set $C$ of size $\lambda$ from $S(\Phi)$.
Then, in the evaluation phase, \thismethod applies a novel \score to evaluate the quality of each solution in $C$ and selects the optimal solution $\alpha$.
In the final optimization phase, \thismethod utilizes \postopt to refine the selected solution $\alpha$, attempting to find a better solution $\alpha^*$, which is then added to the solution set $A$.
When the iteration reaches the termination condition, \ie{} when the computational resources are exhausted or the target AST-coverage $r$ is achieved, \thismethod concludes the iteration process and outputs the current solution set $A$.

\subsection{Sampling Phase}
\label{sec:sampling_phase}

{
\SetAlCapFnt{\small}
\SetAlCapNameFnt{\small}
\SetAlFnt{\small}
\SetKw{Continue}{continue}
\SetKw{Break}{break}

\begin{algorithm} [t]
    \caption{Diversity-aware SMT Algorithm (Diversity\-SMT)}
    \label{algo:diversitysmt}
    
    \KwIn{$\bm{\Phi}$: an SMT formula;
    \newline
    $\bm{A}$: a solution set of $\Phi$\;}
    \KwOut{$\bm{\alpha}$: a solution of $\Phi$\;}

    $\Phi'\leftarrow$ the bit-vector abstraction of $\Phi$\;
    $A'\leftarrow\varnothing$\;

    \ForEach{$\alpha$ in $A$}
    {
        $A'\leftarrow A' \cup \{$the bit-vector abstraction of $\alpha \}$\;
    }

    $B \leftarrow VarBits(\Phi')$\;\label{algoline:cal_distribution_begin}
    
    $dist \leftarrow \varnothing$\;

    \ForEach{$b$ in $B$}
    {
        $dist(b) \leftarrow$ the 0-1 distribution of $b$ in $A'$\;
    }\label{algoline:cal_distribution_end}

    $L'\leftarrow\varnothing$\;

    \Repeat{$L'=\varnothing$}{
        
        $\Phi'\leftarrow\Phi' \cup L$\;\label{algoline:and_L}
        
        $\phi_{b}\leftarrow$ the SAT formula obtained by bit-blasting $\Phi'$\;\label{algoline:bit-blasting}

        $dist' \leftarrow \varnothing$\;
        
        \ForEach{$b$ in $B$}
        {
            $dist'(blasted(b)) \leftarrow dist(b)$\;
        }

        $\alpha_b\leftarrow$\texttt{DiversitySAT}$(\phi_{b}, dist')$\;\label{algoline:call_contextsat}

        $\alpha\leftarrow\Phi'_{\phi_b}(\alpha_b)$\;\label{algoline:get_sol}

        $\alpha, L\leftarrow$\texttt{ArraySolver}$(\alpha)$\;\label{algoline:array_solver}

        \If{$L'\neq\varnothing$}{
            \Continue \;
        }

        $\alpha, L\leftarrow$\texttt{FunctionSolver}$(\alpha)$\;\label{algoline:function_solver}
    }

    \Return{$\alpha$}\;
    
\end{algorithm}
}

As introduced in Section~\ref{sec:overall_design}, during the sampling phase, 
\thismethod employs our \diversitysmt to generate a candidate solution set $C$ of size $\lambda$ 
(a hyperparameter of \thismethod).
The sampled solutions for the current iteration are then selected and optimized from $C$.
The quality of $C$ is crucial to the performance of \thismethod.
If the candidate solutions in $C$ are highly similar to the existing sampled solutions in $A$, 
this iteration contributes little to the overall AST-coverage, 
resulting in redundant solutions and a larger final solution set.
However, efficiently generating diverse solutions remains a challenging task (see Section~\ref{sec:challengs}).
To address this challenge, we design \diversitysmt to generate solutions that are distinct from the existing solution set $A$.

\subsubsection{Lazy SMT Solver}
\label{sec:lazy_smt_solver}

We implement \diversitysmt on a lazy SMT solver, \bitwuzla \cite{Bitwuzla}.
For context,
we briefly describe it.
Unlike
eager SMT solvers, which fully translate the SMT formulas into SAT formulas
before solving,
\bitwuzla first solve the boolean and bit-vector part
and only lazily incorporate the background theory when the
part's solution influences the theory-solving process.
It maintains a bit-vector solver and solvers for each supported theory (\eg{} array and uninterpreted function).
Theory atoms
outside
the bit-vector theory are abstracted as boolean constants \cite{Bitwuzla}.
For example, an equality between two arrays is abstracted as a boolean constant.
As an SMT formula $\Phi$ is transformed into $\Phi'$ through the above abstraction, the solution $\alpha'$ of $\Phi'$ corresponds to the solution $\alpha$ of $\Phi$ and is also called \textbf{the bit-vector abstraction} of $\alpha$, just as \textbf{the bit-vector abstraction} of $\Phi$ refers to $\Phi'$.
If the bit-vector abstraction is satisfiable, the bit-vector solver produces a solution, which is then checked for theory consistency by the solvers for other theories \cite{Bitwuzla}.

\subsubsection{Our Diversity-aware SMT Algorithm}
\label{sec:diversitysmt}

Following the lazy SMT paradigm, for an SMT formula $\Phi$ and its bit-vector abstraction $\Phi'$,
solutions of $\Phi'$ distinct from those in $A'$ (the bit-vector abstractions of $A$)
correspond to solutions of $\Phi$ distinct from those in $A$.
Thus, the key idea of \diversitysmt is to obtain a distinct solution of $\Phi'$ with respect to $A'$
so as to derive a distinct solution of $\Phi$ with respect to $A$.

As outlined in Algorithm~\ref{algo:diversitysmt}, 
\diversitysmt first computes the 0–1 distribution for each bit position in $A'$ 
(Line \ref{algoline:cal_distribution_begin} -– Line \ref{algoline:cal_distribution_end} in Algorithm \ref{algo:diversitysmt}),
which characterizes the frequency of each bit being assigned 0 or 1 among the sampled solutions (\ie{} solutions in $A$).
Then, following the standard lazy SMT solving process, 
\diversitysmt bit-blasts $\Phi'$ into a SAT formula $\phi_b$ 
(Line \ref{algoline:bit-blasting} in Algorithm \ref{algo:diversitysmt}) 
and calls a modified SAT solver (Line \ref{algoline:call_contextsat} in Algorithm \ref{algo:diversitysmt}) 
that takes $\phi_b$ and the 0–1 distribution as input.
This solver alters the variable assignment strategy of a conventional SAT solver: 
for each bit $x$, if 0 occurs more frequently than 1 in the given distribution, 
the solver gives a higher probability to try assigning $x=1$ first, and vice versa.
For a detailed description of the conventional SAT solving process and its variable assignment heuristics, readers may refer to \cite{BieEtAl09,LuoEtAl22}.

From this,
a corresponding solution $\alpha$ to $\Phi'$ is obtained, ensuring that $\alpha$ is distinct from the existing solution set (Line \ref{algoline:get_sol} in Algorithm \ref{algo:diversitysmt}).
Next, $\alpha$ is passed to solvers for other theories (\eg{} arrays and uninterpreted functions) (Line \ref{algoline:array_solver} -- Line \ref{algoline:function_solver} in Algorithm \ref{algo:diversitysmt}).
These
solvers check whether $\alpha$ satisfies the background theory constraints.
If inconsistencies are detected, a conflict $L$ is returned and
integrated
into the solving process (Line \ref{algoline:and_L} in Algorithm \ref{algo:diversitysmt});
otherwise, the solvers refine the solution by resolving the components abstracted in the bit-vector representation.
Through this process, \diversitysmt effectively generates diverse solutions,
mitigating the diversity challenge.

\subsection{Evaluation Phase}
\label{sec:evaluation_phase}

{
\SetAlCapFnt{\small}
\SetAlCapNameFnt{\small}
\SetAlFnt{\small}
\SetKw{Continue}{continue}
\SetKw{Break}{break}

\begin{algorithm} [t]
    \caption{\score (ASTScrFunc)}
    \label{algo:score_function}
    
    \KwIn{$\bm{\Phi}$: an SMT formula;
    \newline
    $\bm{A}$: a solution set of $\Phi$;
    \newline
    $\bm{\alpha}$: a solution of $\Phi$\;}
    \KwOut{the score of $\alpha$\;}

    $score\leftarrow 0$\;

    \ForEach{$(b, v)$ in $ASTBits(\Phi)$} {
        \If{$\left( \forall \beta \in A, \ \beta[\![b]\!] \neq v \right) \land \left( \alpha[\![b]\!] = v \right)$} {
            $score \leftarrow score + 1$\;
        }
    }
    
    \Return{$score$}\;
    
\end{algorithm}
}

As introduced in Section \ref{sec:overall_design}, the goal of
the
evaluation phase is to assess the quality of solutions in $C$ and select the best one.
Designing a reasonable scoring function is therefore crucial.
Since
\thismethod aims to achieve target AST-coverage with minimal solutions,
a straightforward scoring function is to measure the
increase
in the AST-coverage for each solution.
Specifically, we define the following:

\begin{definition}
\label{def:score}
For a solution $\alpha$ and a set $A$,
$scr(A, \alpha)$ denotes
the AST-coverage gain from adding
$\alpha$ to $A$,
\ie{} $scr(A, \alpha) = Coverage(A \cup \{\alpha\}, \Phi) - Coverage(A, \Phi)$.
\end{definition}

From Section \ref{sec:preliminaries},
$Coverage(A, \Phi)$ and $Coverage(A \cup \{\alpha\}, \Phi)$ represent the ratio of $|CoverSet(A, \Phi)|$ and $|CoverSet(A \cup \{\alpha\}, \Phi)|$ to
the total number of all valid AST-bits in the AST of $\Phi$
\ie{} $ValidBits(\Phi)$,
respectively, as defined in Section \ref{sec:preliminaries}.
To compute this score accurately, we must first obtain
$ValidBits(\Phi)$.
However, the total number of AST-bits can be extremely large, and determining the validity of each AST-bit requires invoking the SMT solver, which is highly inefficient.
As introduced in Section \ref{sec:experimental_design},
in the standard SMT sampling benchmarks used in this paper,
the number of AST-bits in
SMT formulas is up to
$3.7\times10^4$, so repeatedly invoking the SMT solver to check validity is unacceptable.

Thus, the naive scoring method in Definition \ref{def:score} is not feasible.
To address this issue, we define $Coverage^*(A, \Phi)$, which represents the ratio of $|CoverSet(A, \Phi)|$ to the total number of all AST-bits (both valid and invalid) for a given solution set $A$ and an SMT formula $\Phi$.

\begin{definition}
For a solution set $A$ and an SMT formula $\Phi$, 
let $CoverSet(A, \Phi)$ denote the set of AST-bits covered by $A$. 
$Coverage^*(A, \Phi)$ is defined as
\[
Coverage^*(A, \Phi) = \frac{|CoverSet(A, \Phi)|}{|ASTBits(\Phi)|}.
\]
\end{definition}

Therefore, \thismethod employs the following \score:
\begin{definition}
\label{def:score2}
For a solution $\alpha$ and a set $A$,
$scr^*(A, \alpha)$ denotes the increment in $Coverage^*$ resulting from adding $\alpha$ to $A$,
\ie{} $scr^*(A, \alpha) = Coverage^*(A \cup \{\alpha\}, \Phi) - Coverage^*(A, \Phi)$.
\end{definition}

Replacing $Coverage(A, \Phi)$ and
$Coverage(A \cup \allowbreak \{\alpha\}, \Phi)$
with $Coverage^*(A, \Phi)$ and $Coverage^*(A \cup \{\alpha\}, \Phi)$ in the score calculation will not affect the relative order of the scores for different solutions.
This is because:
$$
scr(A, \alpha) = \frac{|CoverSet(A \cup \{\alpha\}, \Phi)| - |CoverSet(A, \Phi)|}{ValidBits(\Phi)}
$$
$$
scr^*(A, \alpha) = \frac{|CoverSet(A \cup \{\alpha\}, \Phi)| - |CoverSet(A, \Phi)|}{|ASTBits(\Phi)|}
$$

As $ValidBits(\Phi)$ and $|ASTBits(\Phi)|$ are fixed for a given $\Phi$, score rankings among solutions $\alpha \in C$ stay consistent.
Thus, replacing $scr$ with $scr^{*}$ preserves optimal $\alpha$ selection.

In computing \score as per Definition \ref{def:score2}, it is only necessary to track which AST-bits are newly covered by $\alpha$.
Since an AST-bit covered by
$\alpha$
is guaranteed to be valid, there is no need to invoke the solver to verify its validity.
By replacing the computationally expensive $ValidBits(\Phi)$ with $|ASTBits(\Phi)|$, which can be computed without repeatedly verifying the validity of AST-bits, the efficiency of the evaluation phase is significantly improved.
Through this, \thismethod can efficiently evaluate the quality of a solution and select the best one from the candidate set.

Moreover, as aforementioned, calculating accurate AST-coverage
is challenging.
Therefore, using $Coverage(A, \Phi) \geq r$ as the termination condition for \thismethod would lead to very inefficient performance.
Since $Coverage^*(A, \Phi)$ is much simpler to compute and it holds that $Coverage^*(A, \Phi) \leq Coverage(A, \Phi)$, using $Coverage^*(A, \Phi) \geq r$ as the termination condition is a more efficient and practical choice.

\subsection{Optimization Phase}
\label{sec:optimization_phase}

{
\SetAlCapFnt{\small}
\SetAlCapNameFnt{\small}
\SetAlFnt{\small}
\SetKw{Continue}{continue}
\SetKw{Break}{break}

\begin{algorithm} [t]
    \caption{Post-sampling optimization (PostOpt)}
    \label{algo:optimization_function}
    
    \KwIn{$\bm{\Phi}$: an SMT formula;
    \newline
    $\bm{A}$: a solution set of $\Phi$;
    \newline
    $\bm{\alpha}$: a solution of $\Phi$\;}
    \KwOut{the best solution $\alpha^*$ searched starting from $\alpha$\;}

    $S\leftarrow\{\alpha\}$\;

    \ForEach{$v$ in $V(\Phi)$}
    {
        $\Phi'\leftarrow \Phi\cup\{v \neq \alpha[\![v]\!]\}$\;\label{algoline:add_constraint}

        $\alpha^{*}\leftarrow \texttt{DiversitySMT}$($\Phi'$, $A$)\;\label{algoline:call_diversitysmt}

        \If{\texttt{ASTScrFunc}($\Phi$, $A$, $\alpha^{*}$) $\ge$ \texttt{ASTScrFunc}($\Phi$, $A$, $\alpha$)}
        {
            $S\leftarrow S\cup\{\alpha^{*}\}$\;
        }
    }

    $\alpha^{*}\leftarrow$ the assignment that maximizes \texttt{ASTScrFunc} in $S$\;\label{algoline:choose_mx_score}
    
    \Return{$\alpha^{*}$}\;
    
\end{algorithm}
}

As introduced in Section \ref{sec:challengs}, few existing SMT samplers explicitly optimize for AST-coverage, which results in the scalability challenge.
Therefore, \thismethod takes a targeted approach: after the sampling phase and evaluation phase,
\thismethod identifies a solution $\alpha$, which is the best solution selected from the candidate set rather than from the entire solution space $S(\Phi)$.
This means that $\alpha$ is likely close to the global optimal solution,
although it may not necessarily be the optimal one.
Its neighborhood may still contain superior solutions that have not yet been explored.
Therefore, while
$\alpha$ is a strong candidate, there is still room for improvement.
To address this, \thismethod includes this optimization phase, which further refines the solution $\alpha$ obtained from the previous phases by searching for better solutions in its vicinity.
Considering that local search
has been shown to be
an effective method for solving hard combinatorial optimization problems (\eg{} \cite{CaiEtAl13, CaiEtAl19, HooStu04, LuoEtAl15, LuoEtAl17, MavEtAl17}),
and thus,
we employ a local search method called \postopt to optimize the solutions.
This method refines the selected solution by exploring its neighborhood for better candidates.

In Algorithm \ref{algo:optimization_function}, for a solution $\alpha$, an SMT formula $\Phi$ and a solution set $A$, the \postopt attempts to modify each variable in $V(\Phi)$
one at a time.
It introduces a new constraint
$v \neq \alpha[\![v]\!]$ for each variable $v$ (Line \ref{algoline:add_constraint} in Algorithm \ref{algo:optimization_function}), and then calls \diversitysmt to generate a new solution (Line \ref{algoline:call_diversitysmt} in Algorithm \ref{algo:optimization_function}).
Thereby,
\thismethod gains access to the neighboring solutions around $\alpha$.
Once new solutions are generated, \thismethod evaluates their quality using \score and selects
the highest-scoring solution
and then add it to
$A$ (Line \ref{algoline:choose_mx_score} in Algorithm \ref{algo:optimization_function}).
Through this way, each iteration generate a solution that covers a larger number of AST-bits, resulting in a smaller solution set needed to achieve the targeted AST-coverage.

\section{Experimental Design}
\label{sec:experimental_design}

\subsection{State-of-the-art Competitors}
\label{sec:competitors}

In our experiments, \thismethod is compared against three state-of-the-art SMT sampling competitors, \ie{} \smtsampler \cite{SMTSampler}, \guidedsampler \cite{GuidedSampler}, and \csb \cite{csb}.

\textbf{\smtsampler} \cite{SMTSampler}
samples solutions
from SMT formulas with the theories of bit-vectors, arrays, and uninterpreted functions (\ie{} \smtsampler supports the QF\_BV, QF\_ABV, QF\_AUFBV logics).
Experimental results in the literature \cite{SMTSampler} present that \smtsampler performs well in terms of the number of samples produced and the
AST-coverage.

\textbf{\guidedsampler} \cite{GuidedSampler} is a coverage-guided SMT sampler.
It begins by selecting a random base solution and then applies mutations to generate new solutions from adjacent or unexplored regions of the solution space.
\guidedsampler supports the QF\_BV, QF\_ABV, and QF\_AUFBV logics.

\textbf{\csb} \cite{csb} is a recently-proposed SMT sampler that specifically focuses on the theory of bit-vector (\ie{} \csb only supports the QF\_BV logic).
\csb is developed based on a SAT sampler.
It first uses bit-blasting to convert the SMT formula into a SAT formula,
then applies a SAT sampler to sample solutions, which are converted back into SMT solutions.

The implementations of \smtsampler\footnote{\url{https://github.com/RafaelTupynamba/SMTSampler}}, \guidedsampler\footnote{\url{https://github.com/RafaelTupynamba/GuidedSampler}}, and \csb\footnote{\url{https://github.com/meelgroup/csb}} are all publicly available online.

\subsection{Benchmarks}
\label{sec:benchmarks}

Following the literature that describes \smtsampler \cite{SMTSampler} and \guidedsampler \cite{GuidedSampler},
both of which are competitors of \thismethod, we adopt a public and standard benchmark collection from SMT-LIB \cite{BarEtAl18,SMTLIB}, specifically focusing on the
formulas from
QF\_BV, QF\_ABV, and QF\_AUFBV logics.
These benchmarks have been widely studied
\cite{SMTSampler,GuidedSampler,YaoEtAl20,Bitwuzla,BruEtAl09} and cover a wide range of real-world applications, including software testing, hardware verification, bounded model checking, symbolic execution, and static analysis \cite{SMTSampler,BarEtAl18,SMTLIB}.

Following the benchmark selection process in \cite{SMTSampler} and \cite{GuidedSampler}, we first remove unsatisfiable formulas.
Then, from the remaining benchmarks, we randomly select 213 formulas (the same number of selected formulas as described in the literature \cite{SMTSampler} and \cite{GuidedSampler}), forming a medium-scale benchmark set.
For the benchmarks in the medium-scale benchmark set, the average numbers of total variable bits and AST-bits (as defined in Section \ref{sec:preliminaries}) are 489.3 and 12,744.4, respectively.

As described above, existing studies on SMT sampling \cite{SMTSampler,GuidedSampler} primarily concentrate on medium-scale benchmarks.
Nevertheless, as introduced in Section \ref{sec:introduction}, real-world applications usually involve large-scale SMT formulas, which
are not
effectively handled by existing SMT samplers \cite{SMTSampler,GuidedSampler,csb}.
To bridge this gap, we also construct a large-scale benchmark set.
After filtering the 213 medium-scale benchmarks, we select the top 5\% of benchmarks for each logic based on the total number of variable bits.
Since benchmarks from the QF\_AUFBV logic only have 6 benchmarks in total, we incorporate all of them by exception.
This results in a large-scale benchmark set consisting of 328, 216, and 6 benchmarks from the QF\_BV, QF\_ABV, and QF\_AUFBV logics, respectively, totaling 550 benchmarks.
For the large-scale benchmark set, the average number of total variable bits is 4,104.9,
with the maximum reaching 66,476,
which is significantly larger than this value for the medium-scale benchmark set (\ie{} 489.3);
also, the number of total variable bits for the largest benchmark is 66,476.
For the large-scale benchmark set, the average number of AST-bits is 37,167.9, far exceeding this value for the medium-scale benchmark set (\ie{} 12,744.4).
All adopted benchmarks are publicly available in our repository.\footnote{\url{https://github.com/ShuangyuLyu/PanSampler}\label{url:thismethod_repo}}

\subsection{Research Questions}
\label{sec:research_questions}

In this work, we design our experiments to answer the following research questions (RQs).

\textbf{RQ1:Can \thismethod achieve target AST-coverage on more large-scale benchmarks than its competitors?}

In this RQ, we compare \thismethod with its competitors in terms of their ability to achieve target AST-coverage across the large-scale benchmarks.

\textbf{RQ2: Can \thismethod produce smaller solution sets for the same AST-coverage on large-scale benchmarks?}

In this RQ, we compare the solution set sizes generated by \thismethod and its state-of-the-art competitors on large-scale benchmarks.

\textbf{RQ3: Can \thismethod achieve higher efficiency than its competitors?}

In this RQ, we compare the efficiency of \thismethod and its state-of-the-art competitors on large-scale benchmarks.

\textbf{RQ4: Beyond large-scale benchmarks, how does \thismethod perform on medium-scale benchmarks?}

In this RQ, we compare the performance of \thismethod with its competitors across the medium-scale benchmarks.

\textbf{RQ5: Does each core technique in this work contribute to the performance improvements of \thismethod?}

In this RQ, we analyze the contribution of each core technique to the performance improvement of \thismethod.

\textbf{RQ6: How does the setting of the hyper-parameter impact the practical performance of \thismethod?}

In this RQ, we
investigate how the setting of the hyper-parameter (\ie{} $\lambda$) impacts the performance of \thismethod.

\textbf{RQ7: How does \thismethod perform in real-world software testing?}

In this RQ, we conduct an empirical evaluation on practical subjects, which are collected from real-world software systems, to assess the practical effectiveness of \thismethod in software testing.

\subsection{Experimental Setup}
\label{sec:experimental_setup}

In this work, all experiments were performed on a computing machine
with an AMD EPYC 7763 CPU, 1TB memory, and Ubuntu 20.04.4 LTS.
For \thismethod, we set the hyper-parameter $\lambda$ to 50, and for all its competitors, we used the default configurations recommended by the authors.

In our experiments, the target AST-coverage was set to 80\%, 90\%, 95\%, 98\%, 99\%, and 99.5\%, denoted as `Cov'.
In line with previous studies \cite{MeGASampler}, we normalized AST-coverage
as the ratio of covered AST-bits to the total
AST-bits covered by at least one sampler in the evaluation.
For AST construction, we followed \cite{SMTSampler, MeGASampler}, using the widely recognized \emph{Z3} \cite{MouEtAl08} as the standard.
For each
sampler, we report the number of successful benchmarks where the target AST-coverage is achieved with up to 1,000 solutions and within 1 hour of runtime, denoted as `\texttt{\#}suc'.
The total number of benchmarks where all samplers (\ie{} \thismethod and all its competitors) successfully attain the target AST-coverage is represented as `\texttt{\#}suc.all'.
We also report the average minimum number of solutions required to achieve the target AST-coverage across all successful benchmarks, denoted as `\texttt{\#}sol'.
The average time (in seconds) spent on all benchmarks that successfully achieved the target AST-coverage is denoted as `time'.
Additionally, if a sampler does not support a specific logic, its results are marked by `- -'.
If `\texttt{\#}suc' is 0, both `\texttt{\#}sol' and `time' are also marked by `- -'.
This cutoff time setting is consistent with the experimental setup in \cite{SMTSampler}.
Results in Section \ref{sec:experimental_results} shows the configured time limit and maximum solution count adequately demonstrate samplers' AST-coverage growth trends.
For comparisons using `\texttt{\#}suc', `\texttt{\#}sol', and `time' as metrics, the competing sampler achieving the highest `\texttt{\#}suc' is considered the best, with ties broken by preferring the sampler with the smallest `\texttt{\#}sol' value.
If further ties occur, the sampler with the smallest `time' is preferred.
The best results are highlighted in \textbf{boldface}.

\section{Experimental Results}
\label{sec:experimental_results}

\subsection{RQ1: AST-Coverage Achievement Comparison}

\begin{table}[t]
\centering

    \setlength{\tabcolsep}{4pt}
    \caption{Results of the `\texttt{\#}suc', `\texttt{\#}sol' and `time' (in seconds) metrics obtained by \thismethod and its state-of-the-art competitors on the large-scale benchmark set.}
    \label{tab:vs_sota}
    
    \begin{tabular}{lrrrrrrrr}
        \toprule
        \multirow{3}*{Cov} & \thismethod &  \smtsampler & \guidedsampler & \csb \\
        \cmidrule(lr){2-2}\cmidrule(lr){3-3}\cmidrule(lr){4-4}\cmidrule(lr){5-5}
        & \texttt{\#}suc & \texttt{\#}suc & \texttt{\#}suc & \texttt{\#}suc \\
        & \texttt{\#}sol (time) & \texttt{\#}sol (time) & \texttt{\#}sol (time) & \texttt{\#}sol (time) \\
        
        \midrule
        \multicolumn{5}{l}{\raggedright \emph{Logic: QF\_BV (328 benchmarks)}} \\
        
        \multirow{2}{*}{80.0\%} & \textbf{328} &   294 &   325 &   262 \\   & \textbf{1.2 (37.5)}      &   7.6 (45.1)      &   7.4	(53.2)     &   1.3 (602.2)     \\
        \multirow{2}{*}{90.0\%} & \textbf{328} &   289 &   320 &   217 \\   & \textbf{1.9 (63.0)}      &   29.5 (190.0)    &   50.8 (312.2)    &   2.0 (599.5)     \\
        \multirow{2}{*}{95.0\%} & \textbf{327} &   289 &   320 &   186 \\   & \textbf{2.9 (73.9)}      &   45.1 (304.6)    &   78.0 (525.6)    &   3.2 (445.2)     \\
        \multirow{2}{*}{98.0\%} & \textbf{326} &   288 &   317 &   156 \\   & \textbf{5.0 (105.2)}     &   75.7 (504.0)    &   103.9 (706.7)   &   7.6 (177.9)     \\
        \multirow{2}{*}{99.0\%} & \textbf{324} &   278 &   298 &   147 \\   & \textbf{8.1 (143.6)}     &   140.3 (961.5)   &   179.5 (1206.3)  &   16.2 (187.3)    \\
        \multirow{2}{*}{99.5\%} & \textbf{323} &   160 &   164 &   138 \\   & \textbf{15.6 (244.6)}    &   138.2 (751.6)   &   157.8 (870.9)   &   21.4 (146.4)    \\

        \midrule
        \multicolumn{5}{l}{\raggedright \emph{Logic: QF\_ABV (216 benchmarks)}} \\

        \multirow{2}{*}{80.0\%} & \textbf{216} &   2   &   29  &   - - \\   & \textbf{2.3 (252.0)}     &   227.0 (1366.8)  &   223.6 (1026.2)  &   - - (- -)    \\
        \multirow{2}{*}{90.0\%} & \textbf{210} &   1   &   2   &   - - \\   & \textbf{3.8 (389.6)}     &   363.0 (186.9)   &   182.0 (97.8)    &   - - (- -)    \\
        \multirow{2}{*}{95.0\%} & \textbf{198} &   1   &   1   &   - - \\   & \textbf{6.7 (431.2)}     &   401.0 (201.6)   &   401.0 (192.3)   &   - - (- -)    \\
        \multirow{2}{*}{98.0\%} & \textbf{186} &   0   &   0   &   - - \\   & \textbf{12.4 (729.7)}    &   - -	(- -)       &   - - (- -)       &   - - (- -)   \\
        \multirow{2}{*}{99.0\%} & \textbf{161} &   0   &   0   &   - - \\   & \textbf{17.5 (1001.6)}   &   - -	(- -)       &   - - (- -)       &   - - (- -)   \\
        \multirow{2}{*}{99.5\%} & \textbf{119} &   0   &   0   &   - - \\   & \textbf{23.5 (905.4)}    &   - -	(- -)       &   - - (- -)       &   - - (- -)   \\

        \midrule
        \multicolumn{5}{l}{\raggedright \emph{Logic: QF\_AUFBV (6 benchmarks)}} \\
        \multirow{2}{*}{80.0\%} & \textbf{6}   &   3   &   3  &   - - \\   & \textbf{4.7 (21.8)}      &   315.7 (1364.9)  &   146.7 (1163.0)  &   - - (- -)   \\
        \multirow{2}{*}{90.0\%} & \textbf{6}   &   2   &   2  &   - - \\   & \textbf{7.2 (36.2)}      &   723.0 (2566.9)  &   166.5 (1236.0)  &   - - (- -)   \\
        \multirow{2}{*}{95.0\%} & \textbf{6}   &   0   &   0  &   - - \\   & \textbf{8.8 (44.4)}      &   - - (- -)       &   - - (- -)       &   - - (- -)   \\
        \multirow{2}{*}{98.0\%} & \textbf{6}   &   0   &   0  &   - - \\   & \textbf{13.2 (73.9)}     &   - - (- -)       &   - - (- -)       &   - - (- -)   \\
        \multirow{2}{*}{99.0\%} & \textbf{3}   &   0   &   0  &   - - \\   & \textbf{13.0 (157.5)}    &   - - (- -)       &   - - (- -)       &   - - (- -)   \\
        \multirow{2}{*}{99.5\%} & \textbf{3}   &   0   &   0  &   - - \\   & \textbf{15.3 (263.4)}    &   - - (- -)       &   - - (- -)       &   - - (- -)   \\
        
        \bottomrule
    \end{tabular}
\end{table}

As introduced in Section \ref{sec:challengs}, existing SMT sampler suffer from the scalability challenge,
which this work aims to address.
Therefore, we compare \thismethod against all its competitors
on the set of 550 large-scale benchmarks described in Section \ref{sec:benchmarks}.
The complete experimental results
are available in our repository\footref{url:thismethod_repo}, with overall results in Table \ref{tab:vs_sota}.

As shown in Table \ref{tab:vs_sota},
under various comparative scenarios in terms of the metrics `\texttt{\#}suc'
and `time', \thismethod considerably outperforms its competitors.
\thismethod achieves 99.5\% coverage on 323 QF\_BV benchmarks, compared to
160, 164, and 138
for \smtsampler, \guidedsampler, and \csb, respectively.
For QF\_ABV, \thismethod achieves 99.5\% coverage on 119 of 216 benchmarks, while no competitor achieves this on any benchmark.
For QF\_AUFBV, \thismethod achieves 99.5\% coverage on 3 of 6 benchmarks, whereas no competitor succeeds on any benchmark.
These results demonstrate \thismethod's superior scalability and effectiveness.

According to Section \ref{sec:experimental_design}, the metrics `\texttt{\#}sol' and `time' in Table \ref{tab:vs_sota} are averaged over the benchmarks on which each competing sampler successfully achieves the target AST-coverage.
Since the sets of benchmarks on which each sampler succeeds differ, these two columns in Table \ref{tab:vs_sota} are indicative only.
A more precise comparison of solution set size and runtime is presented in RQ2 and RQ3.

\subsection{RQ2: Solution Set Size Comparison}

\begin{table}[t]

\centering

    \setlength{\tabcolsep}{2pt}
    \caption{Results of the `\texttt{\#}sol' and `time' (in seconds) metrics obtained by \thismethod and its state-of-the-art competitors on the large-scale benchmarks successfully covered by all samplers.}
    \label{tab:vs_sota_size}

    \begin{tabular}{lrrrrrrrrrr}
        
        \toprule
        \multirow{2}*{Cov} & \multirow{2}*{\texttt{\#}suc.all} & \multicolumn{2}{c}{\thismethod} & \multicolumn{2}{c}{\smtsampler} & \multicolumn{2}{c}{\guidedsampler} & \multicolumn{2}{c}{\csb}
        
        \\ \cmidrule(lr){3-4}\cmidrule(lr){5-6}\cmidrule(lr){7-8}\cmidrule(lr){9-10}

        & & \texttt{\#}sol & time & \texttt{\#}sol & time & \texttt{\#}sol & time & \texttt{\#}sol & time \\
        
        \midrule

        80.0\%	& 254	&   \textbf{1.3}	&   \textbf{21.1}   &	6.9	    &   39.3	&   8.2	    &   50.9	&   1.3	    &   568.4   \\
        90.0\%	& 210	&   \textbf{1.7}	&   \textbf{21.6}   &	27.2	&   163.4	&   43.0	&   249.3	&   1.9	    &   614.8   \\
        95.0\%	& 179	&   \textbf{2.5}	&   \textbf{26.3}   &	32.9	&   185.0	&   43.7	&   230.2	&   3.1	    &   452.7   \\
        98.0\%	& 147	&   \textbf{3.8}	&   \textbf{24.3}   &	44.4	&   187.9	&   45.7	&   185.0	&   5.1	    &   158.0   \\
        99.0\%	& 135	&   \textbf{5.7}	&   \textbf{33.7}   &	59.1	&   178.3	&   69.5	&   224.1	&   8.3	    &   68.1    \\
        99.5\%	& 130	&   \textbf{7.4}	&   \textbf{33.1}   &	90.1	&   257.8	&   97.4	&   300.4	&   15.6	&   52.7    \\

    \bottomrule
    \end{tabular}
    
\end{table}

In this subsection, we compare the solution set sizes generated by \thismethod and its competitors.
Table \ref{tab:vs_sota_size} presents the average solution set size and average runtime for each target AST-coverage.
To ensure a fair comparison, these statistics are computed only for the subset of benchmarks on which 
\thismethod and all its competitors successfully achieve the corresponding target AST-coverage, 
so that the comparisons are made over a unified set of benchmarks.

According to Table \ref{tab:vs_sota_size}, for these benchmarks, \thismethod generates significantly smaller solution sets.
Specifically, when the target AST-coverage is 99.5\%, \thismethod requires an average of only 7.4 solutions, whereas the best-performing competitor, \csb, requires 15.6 solutions.
This corresponds to an average reduction of 52.6\% in solution set size.
As discussed in Section \ref{sec:introduction}, this reduction can substantially decrease the time and cost associated with testing in practical scenarios.

\subsection{RQ3: Efficiency Comparison}

In this subsection, we compare the efficiency of \thismethod with its competitors.
Similar to RQ2, to ensure a fair comparison over a unified set of benchmarks, we focus on the subset of benchmarks on which 
\thismethod and all its competitors successfully achieve the corresponding target AST-coverage.
The overall results are presented in Table \ref{tab:vs_sota_size}.

As shown in Table \ref{tab:vs_sota_size}, \thismethod consistently outperforms its competitors across all target AST-coverage levels.
Compared with the best-performing competitor, \thismethod reduces the average runtime by up to
86.8\% at a target AST-coverage of 90.0\%, and by at least
37.2\% at a target AST-coverage of 99.5\%.
These results clearly demonstrate the high efficiency of \thismethod.

\subsection{RQ4: Performance on Medium-Scale Benchmarks}

\begin{table}[t]

\centering

    \setlength{\tabcolsep}{5pt}
    \caption{Results of the `\texttt{\#}suc', `\texttt{\#}sol' and `time' (in seconds) metrics obtained by \thismethod and its state-of-the-art competitors on the medium-scale benchmark set.}
    \label{tab:vs_sota_medium}

    \begin{tabular}{lrrrrrrrrr}
        
        \toprule
        \multirow{3}*{Cov} & \thismethod &  \smtsampler & \guidedsampler & \csb \\
        \cmidrule(lr){2-2}\cmidrule(lr){3-3}\cmidrule(lr){4-4}\cmidrule(lr){5-5}
        & \texttt{\#}suc & \texttt{\#}suc & \texttt{\#}suc & \texttt{\#}suc \\
        & \texttt{\#}sol (time) & \texttt{\#}sol (time) & \texttt{\#}sol (time) & \texttt{\#}sol (time) \\
        
        \midrule
        \multicolumn{5}{l}{\raggedright \emph{Logic: QF\_BV (157 benchmarks)}} \\

        \multirow{2}{*}{80.0\%} & \textbf{157} &   155 &   155 &   144  \\  & \textbf{1.5 (9.1)}   &   17.1 (23.1)     &   25.8 (82.7)     &   1.5 (12.3)  \\
        \multirow{2}{*}{90.0\%} & \textbf{156} &   155 &   154 &   144  \\  & \textbf{2.4 (12.8)}  &   51.7 (40.0)     &   64.9 (133.3)    &   2.5 (32.2)  \\
        \multirow{2}{*}{95.0\%} & \textbf{156} &   153 &   153 &   143  \\  & \textbf{3.9 (17.1)}  &   73.4 (64.7)     &   83.4 (159.1)    &   4.0 (29.0)  \\
        \multirow{2}{*}{98.0\%} & \textbf{155} &   148 &   149 &   143  \\  & \textbf{6.5 (23.2)}  &   101.1 (137.2)   &   106.9 (163.1)   &   7.2 (43.6)  \\
        \multirow{2}{*}{99.0\%} & \textbf{155} &   140 &   136 &   142  \\  & \textbf{15.1 (31.2)} &   117.7 (225.3)   &   118.1 (163.6)   &   12.2 (33.2) \\
        \multirow{2}{*}{99.5\%} & \textbf{155} &   119 &   118 &   142  \\  & \textbf{17.4 (38.7)} &   113.4 (148.7)   &   109.3 (149.6)   &   17.5 (49.2) \\
        
        \midrule
        \multicolumn{5}{l}{\raggedright \emph{Logic: QF\_ABV (56 benchmarks)}} \\
        
        \multirow{2}{*}{80.0\%} & \textbf{56}  &   40  &   43   &   - - \\  & \textbf{2.0 (3.2)}   &   11.5 (2.1)      &   15.6 (3.4)      &    - - (- -)   \\
        \multirow{2}{*}{90.0\%} & \textbf{56}  &   39  &   28   &   - - \\  & \textbf{3.1 (4.5)}   &   28.4 (8.9)      &   27.5 (21.0)     &    - - (- -)   \\
        \multirow{2}{*}{95.0\%} & \textbf{56}  &   39  &   26   &   - - \\  & \textbf{19.8 (7.3)}  &   50.2 (24.1)     &   56.5 (94.2)     &    - - (- -)   \\
        \multirow{2}{*}{98.0\%} & \textbf{50}  &   33  &   16   &   - - \\  & \textbf{30.8 (14.7)} &   103.8 (103.1)   &   207.9 (176.2)   &    - - (- -)   \\
        \multirow{2}{*}{99.0\%} & \textbf{43}  &   26  &   9    &   - - \\  & \textbf{41.0 (22.8)} &   152.3 (171.4)   &   62.9 (111.6)    &    - - (- -)   \\
        \multirow{2}{*}{99.5\%} & \textbf{38}  &   23  &   8    &   - - \\  & \textbf{20.8 (28.7)} &   170.3 (201.2)   &   95.6 (185.4)    &    - - (- -)   \\
                        
    \bottomrule
    \end{tabular}
    
\end{table}

\begin{table}[t]

\centering

    \setlength{\tabcolsep}{2pt}
    \caption{Results of the `\texttt{\#}sol' and `time' (in seconds) metrics obtained by \thismethod and its state-of-the-art competitors on the medium-scale benchmarks successfully covered by all samplers.}
    \label{tab:vs_sota_medium_size}

    \begin{tabular}{lrrrrrrrrrr}
        
        \toprule
        \multirow{2}*{Cov} & \multirow{2}*{\texttt{\#}suc.all} & \multicolumn{2}{c}{\thismethod} & \multicolumn{2}{c}{\smtsampler} & \multicolumn{2}{c}{\guidedsampler} & \multicolumn{2}{c}{\csb}
        
        \\ \cmidrule(lr){3-4}\cmidrule(lr){5-6}\cmidrule(lr){7-8}\cmidrule(lr){9-10}

        & & \texttt{\#}sol & time & \texttt{\#}sol & time & \texttt{\#}sol & time & \texttt{\#}sol & time \\
        
        \midrule

        80.0\%	& 144 &   \textbf{1.5}     &   \textbf{3.6}	    &   15.1	&   9.8	    &   15.9	&   10.6	&   1.5	    &   12.3    \\
        90.0\%	& 142 &   \textbf{2.4}     &   \textbf{5.1}	    &   49.8	&   18.8	&   50.5	&   19.7	&   2.5	    &   32.6    \\
        95.0\%	& 140 &   \textbf{3.8}     &   \textbf{7.1}	    &   72.4	&   36.3	&   68.3	&   36.0	&   3.9	    &   29.6    \\
        98.0\%	& 135 &   \textbf{5.6}     &   \textbf{11.3}	    &   91.8	&   60.2	&   93.4	&   73.2	&   6.5	    &   45.9    \\
        99.0\%	& 122 &   \textbf{9.1}     &   \textbf{14.5}	    &   95.9	&   88.0	&   110.6	&   95.7	&   10.7	&   37.9    \\
        99.5\%	& 102 &   \textbf{11.6}	 &   \textbf{19.1}	    &   98.7	&   83.5	&   101.8	&   116.4	&   15.1	&   37.7    \\

    \bottomrule
    \end{tabular}
    
\end{table}

In addition to evaluating \thismethod on large-scale benchmarks, it is also important to assess its effectiveness on medium-scale benchmarks.
In this subsection, we evaluate \thismethod and its competitors on the medium-scale benchmark set of 213 benchmarks.
As described in Section \ref{sec:experimental_design}, this medium-scale benchmark set is significantly smaller than the previously evaluated large-scale benchmarks
in terms of both the total number of variable bits and the number of AST-bits.

Table \ref{tab:vs_sota_medium} presents the overall experimental results. As shown Table \ref{tab:vs_sota_medium}, \thismethod consistently outperforms its competitors.
Specifically, among 157 QF\_BV and 56 QF\_ABV benchmarks, \thismethod successfully generated a 99.5\% AST-coverage solution set for 155 QF\_BV and 38 QF\_ABV benchmarks.
In contrast, the best-performing competitor achieved this coverage for at most 142 QF\_BV and 23 QF\_ABV benchmarks. Detailed results are available in our public repository\footref{url:thismethod_repo}.

In addition, Table \ref{tab:vs_sota_medium_size} presents the results of \thismethod and its competitors 
on the subset of medium-scale benchmarks on which \thismethod and all its competitors successfully achieve 
the corresponding target AST-coverage.
According to Table \ref{tab:vs_sota_medium_size}, \thismethod also outperforms its competitors 
in both solution set size and efficiency.
Moreover, the advantage becomes more pronounced as the target AST-coverage increases:
when the target AST-coverage reaches 99.5\%,
\thismethod reduces the average solution set size by 23.2\% and the average runtime by 49.3\%.
\subsection{RQ5: Effectiveness of \thismethod's Each Core Technique}

\begin{table}[t]

\centering

    \setlength{\tabcolsep}{4pt}
    \caption{Results of the `\texttt{\#}suc', `\texttt{\#}sol' and `time' (in seconds) metrics obtained by \thismethod and its alternative versions on the large-scale and medium-scale benchmark set.}
    \label{tab:vs_alt}

    \begin{tabular}{lrrrrrrrr}
        \toprule
        \multirow{3}*{Cov} & \thismethod & \altone & \alttwo & \altthree \\
        \cmidrule(lr){2-2}\cmidrule(lr){3-3}\cmidrule(lr){4-4}\cmidrule(lr){5-5}
        & \texttt{\#}suc & \texttt{\#}suc & \texttt{\#}suc & \texttt{\#}suc \\
        & \texttt{\#}sol (time) & \texttt{\#}sol (time) & \texttt{\#}sol (time) & \texttt{\#}sol (time) \\
        
        \midrule
        \multicolumn{5}{l}{\raggedright \emph{Large-scale benchmarks}} \\
        \multirow{2}{*}{80.0\%} & \textbf{550} & 521 & 529 & 523 \\
            & \textbf{1.7 (121.6)} & 1.7 (145.6) & 1.8 (121.8) & 4.2 (62.2) \\

        \multirow{2}{*}{90.0\%} & \textbf{544} & 463 & 472 & 495 \\
            & \textbf{2.7 (188.8)} & 2.8 (261.8) & 2.8 (184.9) & 3.9 (54.6) \\

        \multirow{2}{*}{95.0\%} & \textbf{531} & 369 & 407 & 487 \\
            & \textbf{4.4 (206.8)} & 3.6 (321.4) & 4.6 (148.2) & 4.6 (58.9) \\

        \multirow{2}{*}{98.0\%} & \textbf{518} & 326 & 378 & 484 \\
            & \textbf{7.8 (329.1)} & 5.8 (66.6) & 10.2 (123.5) & 10.9 (180.5) \\

        \multirow{2}{*}{99.0\%} & \textbf{488} & 322 & 363 & 451 \\
            & \textbf{11.3 (426.8)} & 11.1 (113.6) & 18.8 (195.0) & 12.9 (197.4) \\

        \multirow{2}{*}{99.5\%} & \textbf{445} & 318 & 335 & 418 \\
            & \textbf{17.7 (421.4)} & 21.8 (218.9) & 65.9 (547.5) & 22.6 (205.9) \\

        \midrule
        \multicolumn{5}{l}{\raggedright \emph{Medium-scale benchmarks}} \\
        \multirow{2}{*}{80.0\%} & \textbf{213} & 199 & 212 & 212 \\
            & \textbf{1.6 (7.6)} & 9.5 (19.5) & 1.6 (6.8) & 1.9 (6.6) \\

        \multirow{2}{*}{90.0\%} & \textbf{212} & 175 & 210 & 206 \\
            & \textbf{2.6 (10.6)} & 38.9 (20.6) & 2.8 (11.2) & 8.6 (9.7) \\

        \multirow{2}{*}{95.0\%} & \textbf{212} & 147 & 199 & 188 \\
            & \textbf{8.1 (14.5)} & 53.3 (24.7) & 5.0 (13.4) & 3.8 (10.4) \\

        \multirow{2}{*}{98.0\%} & \textbf{205} & 111 & 175 & 168 \\
            & \textbf{12.4 (21.2)} & 43.9 (19.4) & 10.3 (21.8) & 6.8 (19.2) \\

        \multirow{2}{*}{99.0\%} & \textbf{198} & 96 & 162 & 155 \\
            & \textbf{20.7 (29.4)} & 51.8 (55.7) & 14.4 (28.7) & 7.5 (19.4) \\

        \multirow{2}{*}{99.5\%} & \textbf{193} & 90 & 150 & 146 \\
            & \textbf{18.0 (36.7)} & 43.4 (26.5) & 30.0 (58.4) & 10.6 (34.0) \\
        
        \bottomrule
    \end{tabular}

\end{table}

As introduced in Section \ref{sec:our_proposed_algorithm}, \thismethod incorporates three key techniques: \diversitysmt in the sampling phase, \score in the evaluation phase, and \postopt in the optimization phase.
To study the effect of all core techniques, we developed three alternative versions based on \thismethod: \altone, \alttwo, and \altthree.
\altone is the version of \thismethod without the \diversitysmt technique.
In the absence of \diversitysmt, different solutions are obtained by adding additional constraints to the SMT formula.
\alttwo replaces the \score with a more natural way, the Manhattan distance.
In the evaluation phase, \alttwo calculates the Manhattan distance between each solution in $C$ and each solution in $A$, and selects the solution in $C$ that has the largest sum of distances to all solutions in $A$.
\altthree removes the \postopt in \thismethod.
After selecting
a
solution from $C$ in the evaluation phase, it is directly added to $A$ without further optimization.

The complete results are in our repository\footref{url:thismethod_repo} and
Table \ref{tab:vs_alt} presents
the overall performance of \thismethod and all its alternative versions.
It demonstrates that, under various comparative scenarios
\thismethod outperforms all its alternative versions.
These findings indicate that each core technique enhances \thismethod's overall performance.

\subsection{RQ6: Impacts of \thismethod's Hyper-Parameter Settings}

\begin{table}[t]
\centering

\setlength{\tabcolsep}{4pt}
\caption{Results of the `\texttt{\#}suc', `\texttt{\#}sol' and `time' (in seconds) metrics obtained by \thismethod with various settings of $\lambda$ on the large-scale and medium-scale benchmark set.}
\label{tab:vs_lambda}

\begin{tabular}{lrrrrrrrr}
    \toprule
    \multirow{3}*{Cov} & $\lambda = 1$ &  $\lambda = 10$ & $\lambda = 50$ & $\lambda = 100$ \\
    \cmidrule(lr){2-2}\cmidrule(lr){3-3}\cmidrule(lr){4-4}\cmidrule(lr){5-5}
    & \texttt{\#}suc & \texttt{\#}suc & \texttt{\#}suc & \texttt{\#}suc \\
    & \texttt{\#}sol (time) & \texttt{\#}sol (time) & \texttt{\#}sol (time) & \texttt{\#}sol (time) \\
    
    \midrule
    \multicolumn{5}{l}{\raggedright \emph{Large-scale benchmarks}} \\
    \multirow{2}{*}{80.0\%} & 533 & 545 & \textbf{550} & 522 \\
        & 1.8 (121.8) & 1.7 (93.0) & \textbf{1.7 (121.6)} & 1.7 (144.9) \\

    \multirow{2}{*}{90.0\%} & 482 & 534 & \textbf{544} & 466 \\
        & 2.9 (207.1) & 2.7 (134.1) & \textbf{2.7 (188.8)} & 2.8 (278.6) \\

    \multirow{2}{*}{95.0\%} & 402 & 520 & \textbf{531} & 380 \\
        & 4.5 (257.8) & 4.3 (174.1) & \textbf{4.4 (206.8)} & 3.5 (276.1) \\

    \multirow{2}{*}{98.0\%} & 358 & 504 & \textbf{518} & 345 \\
        & 7.5 (107.3) & 7.7 (267.1) & \textbf{7.8 (329.1)} & 5.4 (133.6) \\

    \multirow{2}{*}{99.0\%} & 348 & 473 & \textbf{488} & 340 \\
        & 11.1 (183.1) & 11.3 (312.8) & \textbf{11.3 (426.8)} & 8.3 (194.1) \\

    \multirow{2}{*}{99.5\%} & 331 & 443 & \textbf{445} & 334 \\
        & 18.7 (172.5) & 20.0 (307.7) & \textbf{17.7 (421.4)} & 15.4 (275.5) \\

    \midrule
    \multicolumn{5}{l}{\raggedright \emph{Medium-scale benchmarks}} \\
    \multirow{2}{*}{80.0\%} & 211 & 213 & \textbf{213} & 210 \\
        & 2.1 (2.8) & 1.7 (5.4) & \textbf{1.6 (7.6)} & 1.6 (11.9) \\

    \multirow{2}{*}{90.0\%} & 210 & 212 & \textbf{212} & 208 \\
        & 12.5 (5.7) & 3.7 (7.4) & \textbf{2.6 (10.6)} & 2.5 (17.6) \\

    \multirow{2}{*}{95.0\%} & 190 & \textbf{212} & 212 & 195 \\
        & 23.1 (10.2) & \textbf{7.7 (10.7)} & 8.1 (14.5) & 5.6 (21.5) \\

    \multirow{2}{*}{98.0\%} & 178 & \textbf{206} & 205 & 176 \\
        & 50.0 (43.8) & \textbf{11.9 (25.6)} & 12.4 (21.2) & 7.9 (27.9) \\

    \multirow{2}{*}{99.0\%} & 160 & \textbf{198} & 198 & 164 \\
        & 65.6 (64.6) & \textbf{18.4 (36.2)} & 20.7 (29.4) & 9.9 (37.9) \\

    \multirow{2}{*}{99.5\%} & 149 & 193 & \textbf{193} & 158 \\
        & 70.5 (58.5) & 25.9 (44.3) & \textbf{18.0 (36.7)} & 12.2 (43.1) \\
    \bottomrule
\end{tabular}

\end{table}

\thismethod has one hyper-parameter, $\lambda$, which represents the size of the candidate set $C$.
Table \ref{tab:vs_lambda} reports the performance of \thismethod with different values of $\lambda$ (\ie{} $\lambda = 1, 10, 50, 100$).
A larger $\lambda$ allows \thismethod to sample more solutions during each iteration's sampling phase, enabling a broader exploration of the solution space.
This increases the likelihood of identifying solutions that yield a greater AST-coverage increment in the current iteration,
leading to a smaller solution set.
However, as the candidate set size increases, the efficiency of \thismethod decreases.
As shown in Table \ref{tab:vs_lambda},
When $\lambda$ is too small (\eg{} $\lambda = 1$), \thismethod requires more solutions to achieve the target coverage, resulting in lower `\texttt{\#}suc', higher `\texttt{\#}sol', and increased runtime due to
the additional iterations needed to reach the target AST-coverage.
Conversely, when $\lambda$ is too large (\eg{} $\lambda = 100$), the efficiency of \thismethod decreases, reducing the number of benchmarks that reach the target coverage within the cutoff limit.
Based on these results, we select $\lambda = 50$ as the default setting for \thismethod, as it achieves the best overall performance.

\subsection{RQ7: Performance in Real-World Software Testing}
\label{sec:RQ7}

\begin{table}[t]
\centering

    \setlength{\tabcolsep}{4pt}
    \caption{Fault detection performance (`\texttt{\#}suc', `\texttt{\#}sol', `time') of \thismethod and its state-of-the-art competitors on the 9 real-world software systems.}
    \label{tab:fdr}
    
    \begin{tabular}{lrrrrrrrr}
        \toprule
        \multirow{3}*{FDR} & \thismethod &  \smtsampler & \guidedsampler & \csb \\
        \cmidrule(lr){2-2}\cmidrule(lr){3-3}\cmidrule(lr){4-4}\cmidrule(lr){5-5}
        & \texttt{\#}suc & \texttt{\#}suc & \texttt{\#}suc & \texttt{\#}suc \\
        & \texttt{\#}sol (time) & \texttt{\#}sol (time) & \texttt{\#}sol (time) & \texttt{\#}sol (time) \\
        
        \midrule

\multirow{2}{*}{70.0\%}	&   \textbf{9}	&   4	&   4	&   1	\\  &   \textbf{7.0 (0.2)}	    &   11.8 (0.1)  &   	11.8 (0.1)  &   	4.0 (0.1) \\
\multirow{2}{*}{80.0\%}	&   \textbf{8}	&   4	&   4	&   1	\\  &   \textbf{10.2 (0.3)}	    &   17.5 (0.1)  &   	35.2 (0.1)  &   	4.0 (0.1) \\
\multirow{2}{*}{90.0\%}	&   \textbf{6}	&   4	&   4	&   0	\\  &   \textbf{16.3 (0.5)}	    &   135.8 (0.3) &   	95.2 (0.3)  &   	- - (- -) \\
\multirow{2}{*}{95.0\%}	&   \textbf{5}	&   3	&   3	&   0	\\  &   \textbf{113.8 (1.3)}	&   226.0 (0.5) &   	142.7 (0.4) &   	- - (- -) \\
\multirow{2}{*}{98.0\%}	&   \textbf{5}	&   3	&   3	&   0	\\  &   \textbf{149.6 (2.4)}	&   230.0 (0.5) &   	144.7 (0.4) &   	- - (- -) \\
\multirow{2}{*}{99.0\%}	&   \textbf{4}	&   3	&   3	&   0	\\  &   \textbf{73.5 (2.3)}	    &   230.0 (0.5) &   	144.7 (0.4) &   	- - (- -) \\
\multirow{2}{*}{99.5\%}	&   \textbf{4}	&   3	&   3	&   0	\\  &   \textbf{73.5 (2.3)}	    &   230.0 (0.5) &   	144.7 (0.4) &   	- - (- -) \\
        
        \bottomrule
    \end{tabular}
\end{table}

\begin{table}[t]

\centering

    \setlength{\tabcolsep}{6pt}
    \caption{Results of the `\texttt{\#}sol' and `time' (in seconds) metrics on the subjects successfully attained by all competing samplers.}
    \label{tab:fdr_size}

    \begin{tabular}{lrrrrrrrr}
        
        \toprule
        \multirow{2}*{FDR} & \multirow{2}*{\texttt{\#}suc.all} & \multicolumn{2}{c}{\thismethod} & \multicolumn{2}{c}{\smtsampler} & \multicolumn{2}{c}{\guidedsampler} 
        
        \\ \cmidrule(lr){3-4}\cmidrule(lr){5-6}\cmidrule(lr){7-8}

        & & \texttt{\#}sol & time & \texttt{\#}sol & time & \texttt{\#}sol & time \\
        
        \midrule
        
70.0\%	& 4	&   \textbf{4.5}     &      0.2	&   11.8	&   0.1	&   11.8	&   0.1 \\
80.0\%	& 4	&   \textbf{11.8}    &   	0.4	&   17.5	&   0.1	&   35.3	&   0.1 \\
90.0\%	& 4	&   \textbf{22.5}    &   	0.7	&   135.8	&   0.3	&   95.3	&   0.3 \\
95.0\%	& 3	&   \textbf{36.7}    &   	1.2	&   226.0	&   0.5	&   142.7	&   0.3 \\
98.0\%	& 3	&   \textbf{96.3}    &   	3.0	&   230.0	&   0.5	&   144.7	&   0.3 \\
99.0\%	& 3	&   \textbf{96.3}    &   	3.0	&   230.0	&   0.5	&   144.7	&   0.3 \\
99.5\%	& 3	&   \textbf{96.3}    &   	3.0	&   230.0	&   0.5	&   144.7	&   0.3 \\

    \bottomrule
    \end{tabular}
    
\end{table}

\subsubsection{\textbf{Subjects and SMT Encoding}}

As described in Section \ref{sec:experimental_design}, the benchmarks utilized in RQ1–RQ6 already encompass a wide range of real-world applications.
The experimental results reported in RQ1–RQ4 demonstrated that \thismethod consistently exhibits a stronger capability to reach target AST-coverage, while requiring fewer solutions than its competitors to achieve the same AST-coverage level.
In this RQ, we conduct a further empirical evaluation to specifically assess the practical effectiveness of \thismethod in the context of real-world software testing.

To achieve this, we employ 9 real-world software systems as our subjects, all of which were originally collected and utilized by a recent empirical study \cite{WuEtAl20}.
These systems are highly configurable, exposing various configuration options that allow for customization.
However, not all configurations (\ie{} combinations of option values) are valid due to complex inter-option constraints.
For example, a constraint from the Linux kernel dictates that the \texttt{X86\_BIGSMP} option,
designed to overcome the 8-CPU limit of standard 32-bit Symmetric Multiprocessing (SMP),
can only be enabled if the architecture option is \texttt{X86\_32} (\ie{} \texttt{X86\_BIGSMP} $\implies$ \texttt{X86\_32}).
According to \cite{WuEtAl20}, some faults in these systems are triggered only when specific combinations of option values are present.
For instance, a known fault of the Linux kernel is only triggered when the \texttt{X86} architecture option and the \texttt{ACPI\_WMI} option (a power management feature) are simultaneously enabled in a configuration.
Effective testing thus requires covering a diverse set of valid configurations to find these fault-triggering interactions, which remains a significant challenge \cite{WuEtAl20,LuoEtAl21b,LuoEtAl22}.
A promising solution is to encode the configuration options and their constraints as an SMT formula.
SMT sampling can then be applied, where each solution generated by the sampler corresponds to a valid configuration, \ie{} a valid test case.
The following example illustrates how a highly configurable software system can be encoded as an SMT formula and tested via SMT sampling.

\begin{example}

Consider a software system with three options:
$l$, a Boolean option;
$m$, a 32-bit unsigned integer-valued option;
and \texttt{mode}, an option with four possible values.  
These options obey the following constraints:
1) if $m = 3$ then $l$ must be true;  
2) if \texttt{mode} takes its fourth possible value then $m > 10$.
Some faults occur only when specific combinations of options are present, for example, when $m = 20$ and $l = \text{false}$.

When encoding this software system as an SMT formula, $l$ can be represented as a Boolean variable, $m$ as a $BV_{32}$ variable,
% and \texttt{mode} as a $BV_2$ variable.
and $mode$ as a $BV_2$ variable, where the four possible values are encoded as \texttt{\#b00}, \texttt{\#b01}, \texttt{\#b10}, and \texttt{\#b11},
corresponding to the first, second, third, and fourth possible values, respectively.
These constraints can be encoded as the following SMT-LIB assertions:
1) \texttt{(assert (=> (= m \#x00000003) l))}
2) \texttt{(assert (=> (= mode \#b11) (bvugt m \#x0000000A)))}

SMT sampling then generates diverse valid configurations that satisfy all these constraints, with each solution corresponding to one test case.  
For instance, a solution such as $(m = \texttt{\#x00000014}, l = \text{false}, mode = \texttt{\#b11})$ may trigger the aforementioned fault, while another configuration $(m = \texttt{\#x00000003}, l = \text{true}, mode = \texttt{\#b00})$ does not.

\end{example}

Accordingly, we encoded these 9 software systems from \cite{WuEtAl20} into SMT formulas, accompanied by an associated list of known faults,
where each fault is annotated with the specific combinations of configuration options that trigger it.
All subjects are publicly available in our repository\footref{url:thismethod_repo}.

\subsubsection{\textbf{Setup and Results of Empirical Study}}

In this RQ, we compare the software testing performance of \thismethod against all its competitors on these 9 real-world systems.
We utilize two primary evaluation metrics.
To assess effectiveness, we use the fault detection rate (FDR).
For a given set of solutions (\ie{} test cases) generated by a sampler, the FDR is calculated as the ratio of the number of unique known faults triggered by at least one of these solutions to the total number of known faults in the subject software system.
To assess efficiency, we record the minimum number of solutions a sampler needs generate to reach a target FDR level.
A lower number of solutions indicates higher efficiency, as it implies less  testing effort is required.

Table \ref{tab:fdr} reports the overall results; detailed experimental data is available in our repository\footref{url:thismethod_repo}.
In this evaluation, we set the target FDR to: 70\%, 80\%, 90\%, 95\%, 98\%, 99\%, and 99.5\%.
All other experimental settings are identical to those described in Section \ref{sec:experimental_design}.
Similar to the setup in Section \ref{sec:experimental_design}, for any given sampler, `\texttt{\#}suc' denotes the number of subjects where it successfully reached the target FDR and
`\texttt{\#}suc.all' represents the total number of subjects where all competing samplers successfully attained the target FDR.
The average minimum number of solutions required to achieve the target FDR and the average time (in seconds) across all successful subjects are denoted as `\texttt{\#}sol' and `time'.

As shown in Table \ref{tab:fdr}, \thismethod exhibits a markedly stronger fault detection capability.
It consistently achieves a higher `\texttt{\#}suc' value than all its competitors across all target FDR levels.
Notably, \thismethod successfully detects at least 70\% of the faults for all 9 subjects, whereas its competitors can, at best, detect 70\% of the faults for only 4 subjects.

When focusing on the subjects where all competing samplers succeeded
for a fair comparison of the required solutions for the target FDR level,
the results are presented in Table \ref{tab:fdr_size}.
Note that \csb is omitted from this part of the analysis due to its markedly smaller `\texttt{\#}suc' value.
According to Table \ref{tab:fdr_size}, \thismethod generates markedly fewer solutions to achieve the same target FDR at all target FDR levels.
In the best-case scenario, at a target FDR of 90\%, it reduces the number of required solutions by 76.4\%, representing a substantial improvement in practical testing efficiency.
Even in the worst case, at a target FDR of 80\%, it still achieves a 32.6\% reduction.

\subsection{Threats to Validity}

\subsubsection{Representativeness of the Benchmarks}
As introduced in Section \ref{sec:experimental_design}, the benchmarks we used cover a wide range of problems and scales.
These benchmarks have been extensively studied in several previous works \cite{SMTSampler, GuidedSampler, YaoEtAl20, Bitwuzla, BruEtAl09}.
In addition, \thismethod's competitors, \smtsampler and \guidedsampler, also use the same benchmarks for evaluation.
Our experiments involved testing 550 large-scale benchmarks and 213 medium-scale benchmarks, while \smtsampler and \guidedsampler only tested 213 medium-scale benchmarks \cite{SMTSampler, GuidedSampler}.
This more extensive evaluation alleviates this potential threat.

\subsubsection{Scope of Logic Support}
SMT includes a wide range of theories, such as bit-vectors, integer arithmetic, strings, and others.
\thismethod currently supports the QF\_BV, QF\_ABV, and QF\_AUFBV logics.
However, the framework for SMT sampling proposed by \thismethod is not limited to these specific logics.
As the theory of fixed-size bit-vectors
and its extended logics are widely used in various applications \cite{BarEtAl16, KovEtAl16}, we initially implemented \thismethod for bit-vector and its extensions,
and we plan to extend its capabilities to other SMT theories in the future.

\subsubsection{Practical Relevance}
This paper focuses on SMT sampling algorithms, which have effective applications in software testing and verification, including symbolic execution and random testing \cite{QuickSampler,JFSampler}.
To address the potential threat regarding practical relevance, we conducted an empirical evaluation in Section \ref{sec:RQ7} on 9 real-world software systems \cite{WuEtAl20}.
The experimental results demonstrated that \thismethod is beneficial for software testing, which alleviates this potential threat.

\section{Related Work}
\label{sec:related_work}

Constraint sampling techniques, including SAT sampling and SMT sampling, have been extensively applied in software and hardware verification and testing \cite{NavEtAl06,ClaEtAl03,GurEtAl15,HeiEtAl13,ChaEtAl15,SMTSampler,GuidedSampler,DenEtAl09,QuickSampler,LuoEtAl22,NanEtAl11}.
Numerous studies have focused on SAT sampling.
For instance, samplers based on Markov Chain Monte Carlo (MCMC) \cite{KitEtAL07, KitEtAL10}, universal hashing \cite{ChaEtAl14, ChaEtAl15}, and model counting \cite{ChaEtAl13,ChaEtAl16} have been explored.
Other approaches include modifications to the internal search heuristics of constraint solvers \cite{NadEtAl11,GolEtAl21} and epoch-based methods \cite{QuickSampler}.

However, while SAT sampling has been widely studied, research on SMT sampling is relatively scarce.
One recent sampler, \csb \cite{csb}, eagerly converts SMT formulas into SAT formulas for sampling, which leads to a loss of higher-level structure in the formulas \cite{SMTSampler}, resulting in a diversity reduction of the sampled solutions.
Other SMT samplers \cite{SMTSampler,GuidedSampler,MeGASampler} employ techniques that generate a large number of similar solutions based on a single solution.
While this improves efficiency, it diminishes the diversity of the sampled solutions.
Furthermore, the coverage-guided fuzzing sampler \cite{JFSampler} utilizes abstract syntax trees (ASTs) to direct the sampling process, thereby enhancing AST-coverage.
However, this method relies on an SMT solver designed specifically for floating-point constraints, limiting its applicability to other theories.
Consequently, as discussed in Section \ref{sec:introduction}, all these existing SMT samplers face significant scalability challenges.
When handling large-scale formulas, they struggle to generate small-sized solution sets that effectively achieve the target AST coverage, which undermines the efficiency of verification and testing in practice.

Thus, this work proposes \thismethod, an SMT sampler that effectively mitigates the scalability challenge.
Implemented for QF\_AUFBV and its sub-logics QF\_ABV and QF\_BV, \thismethod directly optimizes for AST-coverage during the sampling process, resulting in small-sized solution sets.
Furthermore, we presents three novel
techniques \ie{} \diversitysmt,
\score and \postopt to
enhance the performance of \thismethod.
When sampling large-scale SMT formulas, it
demonstrates stronger ability to reach high target AST-coverage and generates smaller solution sets, significantly improving the efficiency and effectiveness of practical verification and testing.
Our experiments on practical subjects from real-world software systems further confirm these advantages.

\section{Conclusion}
\label{sec:conclusion}

In this work, we introduce a novel SMT sampler, dubbed \thismethod, designed to achieve high target AST-coverage with as few solutions as possible, for the theories of bit-vectors, arrays, and uninterpreted functions.
\thismethod alleviates the scalability challenge faced by existing SMT samplers.
Extensive experiments across a wide range of public benchmarks demonstrate that \thismethod exhibits a substantially stronger ability to reach high target AST-coverage, particularly on large-scale SMT formulas.
Moreover, when the target AST-coverage is reached, \thismethod produces smaller solution sets.
The former contributes to improving the effectiveness of practical software testing and hardware verification,
while the latter enhances their efficiency.
In addition, our empirical evaluation on practical subjects, which are collected from real-world software systems, confirmed that \thismethod enhances practical software testing, achieving higher fault detection capability and reducing the number of required test cases from 32.6\% to 76.4\% to reach the same fault detection effectiveness.
Overall, \thismethod advances the state of the art in SMT sampling and provides practical benefits in fields such as software testing and hardware verification.

\section{Data Availability}
\label{sec:data_availability}

The implementation of our
proposed
\thismethod, all adopted benchmarks used in this work, as well as detailed experimental results are publicly available in our repository:
\textbf{\url{https://github.com/ShuangyuLyu/PanSampler}}.

\section*{Acknowledgment}
This work was supported in part by the National Key Research and Development Program of China under Grant 2023YFB3307503, in part by the National Natural Science Foundation of China under Grants 62202025 and 62302528, in part by Beijing Natural Science Foundation under Grant L241050, in part by the Young Elite Scientist Sponsorship Program by CAST under Grant YESS20230566, in part by CCF-Huawei Populus Grove Fund under Grant CCF-HuaweiFM2024005,
and in part by the Fundamental Research Fund Project of Beihang University.

\bibliographystyle{IEEEtran}
\bibliography{main}

\end{document}